\documentstyle[twoside,fleqn,espcrc2,epsfig]{article}
\newcommand{\ec}{\end{center}}
\newcommand{\AmS}{{\protect\the\textfont2
A\kern-.1667em\lower.5ex\hbox{M}\kern-.125emS}}

\begin{document}

\title{Bottomonia hadroproduction
\thanks{Work partially supported by CICYT under contract AEN99-0692}}
\author{J.L. Domenech Garret$^{a}$ and M. A. Sanchis-Lozano$^{b,c}$\thanks{E-mail:domenech@evalo1.ific.uv.es  mas@evalo1.ific.uv.es}
\vspace{0.4cm}\\
(a) Departamento de F{\'\i}sica At\'omica, Molecular  y Nuclear \\
\vspace{0.1cm}
(b) Departamento de F\'{\i}sica Te\'orica \\
\vspace{0.1cm}
(c) Instituto de F\'{\i}sica Corpuscular (IFIC)\\
Centro Mixto Universitat de Val\`encia-CSIC \\
Dr. Moliner 50, E-46100 Burjassot, Valencia (Spain)}
%\begin{document}
\begin{abstract}

We analyze Tevatron data of bottomonium  hadroproduction in the framework of the colour-octet model (COM) implemented in the event generator PYTHIA \cite{pythia} using CTEQ4L PDF taking into account initial-state radiation of gluons and  Altarelli-Parisi evolution of final-state gluons. We obtain new values for the colour-octet matrix elements (Me's) relevant to this production process for the $\Upsilon(nS)$ family (n=1,2,3), finding that $^1S_0^{(8)}+^3P_J^{(8)}$ contributions are not needed in the fit. We show the different contributions to $\Upsilon(1S)$ production at Tevatron for $P_T>8$ GeV, comparing them with CDF data. Remarkably we find a quite small contribution (compatible with zero) from feeddown of $\chi_{bJ}$ states produced through the colour-octet mechanism: $\Upsilon(1S)$ indirect production via $\chi_{bJ}$ decays should be mainly ascribed to the colour-singlet model. Finally we extrapolate to LHC energies to predict $\Upsilon(nS)$ production rates.
\end{abstract}
\vspace{0.1in}
\maketitle

\def\slash{\!\!\!\!/}
\def\slash{\!\!\!/}

\section{HIGHER ORDER QCD EFFECTS; GENERATING $\Upsilon(nS)$}

In a previous work \cite{nos00}, we extracted colour-octet Me's for $\Upsilon(1S)$ hadroproduction using CTEQ2L parton distribution funcion (PDF) \cite{cteq2l}; in this work we extend this study to the upper resonances $\Upsilon(2S)$ and $\Upsilon(3S)$, using the improved CTEQ4L PDF \cite{cteq4l}. We have implemented COM in the event generator PYTHIA with initial-state radiation of gluons \cite{mas0,mas1} and  Altarelli-Parisi evolution of final-state gluons in the same way that earlier.

First we repeat our fit of the $\Upsilon(1S)$ differential cross section with CTEQ4L PDF. We generated separately each source: direct $\Upsilon(1S)$ production from both Colour Singlet Model (CSM) and COM. Indirect source comes from  electromagnetic decays of $\chi_{bJ}(1P)$ and  $\chi_{bJ}(2P)$ and contributions from strong and electromagnetic decays of $\Upsilon(2S)$ and $\Upsilon(3S)$; the $M_5=5{\times}\biggl(\frac{<O_8(^3P_J)>}{m_b^2}+\frac{<O_8(^1S_0)>}{5}\biggr)$ combination is also taken into account (although later we shall see that actually this contribution is not needed). For the CSM Me's we take the values from the Buchmuller and Tye QCD potencial \cite{eichten}. We make use of the Heavy Quark Spin Symmetry in order to relate the different matrix elements.

 We set the $\Upsilon(1S)$ mass as $9.46$ GeV (the mass of 
the coloured intermediate state was set as $2M_b=9.76$ GeV, the difference
is accounted for by soft gluon emission) whereas for $\chi_{bJ}(nP)$  
 we took a weighted mean of their real masses \cite{pdg}. 
 For $\Upsilon(2S)$ we consider direct CSM contribution  and the decay from $\Upsilon(3S)$ and $\chi_{bJ}(2P)$. COM contribution also is taken into account.  For $\chi_{bJ}(2P)$ we set its real mass. Regarding the $\Upsilon(3S)$ state we consider only direct production for both CSM and COM contributions; we set its real mass, excluding any $\chi_{bJ}(3P)$ decays. 

We have included the higher order QCD effects on the partonic  cross sections caused by:

\begin{itemize}

\item Intrinsic Fermi motion of partons inside the hadrons. This non perturbative effect is relevant at small $P_t$ values.
%\newpage
%.\newpage
%.\newpage
\item Multiple emission of gluons in the initial state. This perturbative contribution  is dynamically generated via gluon radiation implemented in the event generator PYTHIA by means of a parton shower algorithm. This effect overshadows the former at high $P_t$ values.       

\item Altarelli-Parisi (AP) evolution of the splitting gluon in the $ gg{\rightarrow} b\-\overline{b}g $ channel. Actually we do not  generate the virtual gluon ($g^*$) that splits into $b\-\overline{b}$ pairs; final hadronization into a ($Q\overline{Q}$) bound state is taken into account by means of the colour-octet matrix elements multiplying the respective short-distance cross sections. Nevertheless, it is reasonable to assume that, on the average, the virtual $g^{\ast}$ should evolve at high $p_T$ similarly to the other final-state gluon - which actually is evolved by the PYTHIA machinery. We used this fact to simulate the (expected) evolution of the (ungenerated)  $g^{\ast}$ whose momentum was assumed to coincide with that of the resonance (neglecting the effect of emission/absorption of soft gluons by the intermediate coloured state bleeding off colour \cite{mas1}).

\end{itemize}                        

\section{RESULTS:}

In table 1 we present the results of our fit to CDF data \cite{fermi1}. Table 2 shows our previous result using CTEQ2L (both in $10^{-3}$ $GeV^3$ units); errors are statistical only. Figures 1 and 2 show the theoretical curves obtained for $\Upsilon(1S)$ and $\Upsilon(3S)$ respectively. The statistical $\chi^2/Ndf$ are quite good, especially for $\Upsilon(3S)$. Those values are consistent with NRQCD velocity scaling rules. Let us remark that due to the $p_t$ cut-off parameter set in the generation, only those experimental points  for $P_t > 1$ GeV were used in the fit.  \\   

\ \ \underline{Table1. COM parameters (CTEQ4L):}
\begin{eqnarray}
<O_8^{\Upsilon(1S)}(^3S_1)>{\mid}_{tot} & = & 80 \pm 28 \nonumber \\ 
<O_8^{\Upsilon(2S)}(^3S_1)>{\mid}_{tot} & = & 77 \pm 42 \nonumber \\
<O_8^{\Upsilon(3S)}(^3S_1)>{\mid}_{tot} & = & 90 \pm 32 \nonumber \\
\nonumber
\end{eqnarray}

\ \ \underline{Table2. COM parameters (CTEQ2L):}\\
\begin{eqnarray}
<O_8^{\Upsilon(1S)}(^3S_1)>{\mid}_{tot} & = & 139 \pm 31 \nonumber \\ 
<O_8^{\Upsilon(2S)}(^3S_1)>{\mid}_{tot} & = & 80 \pm 44 \nonumber \\
<O_8^{\Upsilon(3S)}(^3S_1)>{\mid}_{tot} & = & 75 \pm 27  \nonumber \\     
\nonumber
\end{eqnarray}

\begin{figure}[htb!]
\centerline{
\epsfig{figure=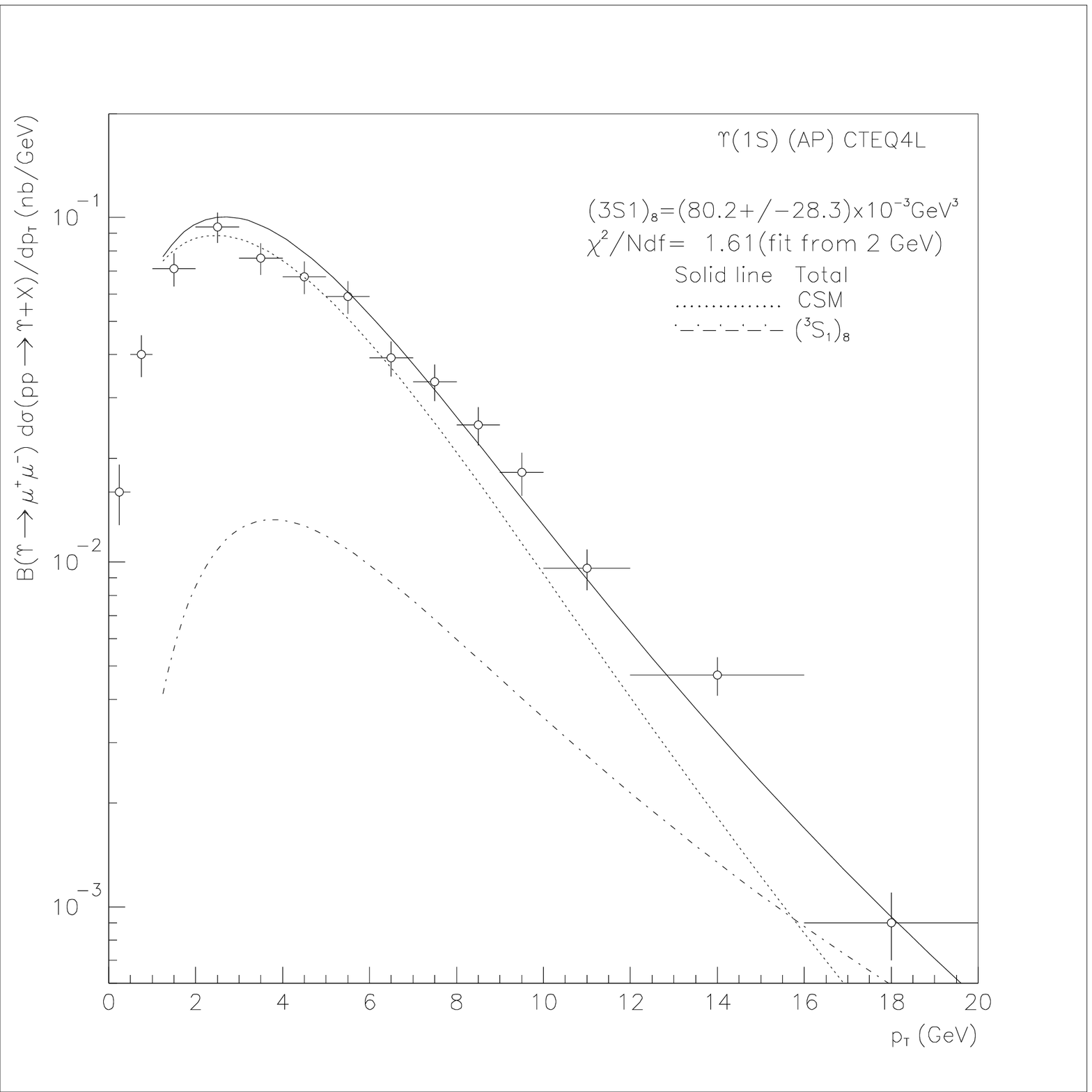,height=7.5cm,width=7.5cm}}
\caption{Theoretical curve obtained from the fit using CTEQ4L PDF in $\Upsilon (1S)$ with AP evolution.}
\end{figure}

\begin{figure}[htb!]
\centerline{
\epsfig{figure=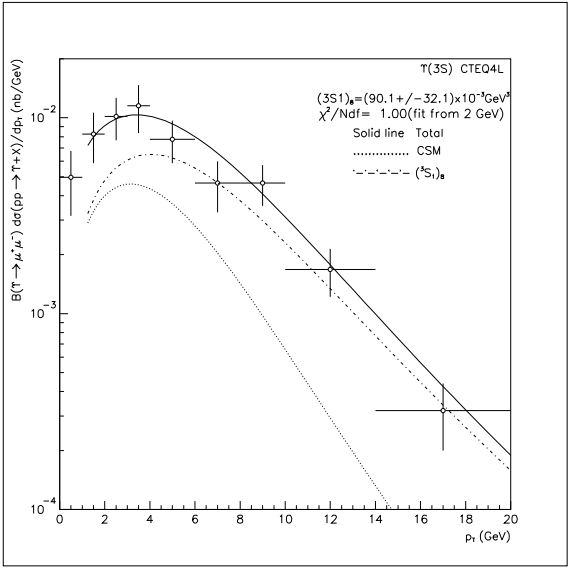,height=7.5cm,width=7.5cm}}
\caption{Theoretical curve obtained from the fit using CTEQ4L PDF in $\Upsilon (3S)$ with AP evolution.}
\end{figure}

As a check of our generation we reproduce the relative fractions of the different contributions to $\Upsilon(1S)$ production obtained by the CDF Collaboration \cite{fermi2}. In table 3 we quote the results from CDF data at $P_T > 8$ GeV; in table 4 the results from our fit in the same case. The agreement is quite good taking into account the errors. If we compare experimental data versus generation, we can see that the $^3S_1^{(8)}$ contribution goes mainly to {\em direct} $\Upsilon(1S)$ production and therefore the $\chi_{bJ}(nP)$ channel would be mainly saturated by the CSM. As a consequence the $\Upsilon(1S)$ COM matrix element should be mainly direct.
%\newpage
%.\newpage
%.\newpage

%\begin{figure}[htb!]
%\centerline{
%\epsfig{figure=upsi3s18_c4l.eps,height=7.5cm,width=7.5cm}}
%\caption{Theoretical curve obtained from the fit using CTEQ4L PDF in $\Upsilon (3S)$ with AP evolution.}
%\end{figure}

Fitting the $\Upsilon(3S)$ differential distribution we assumed only direct production (i.e. no feeddown from higher resonances) and hence the COM parameter must be entirely ascribed to direct contribution. If we compare with the previous values obtained by Cho and Leibovich \cite{cho} for direct production: 5.9; 4.1; 4.1 (in $10^{-3}$ $GeV^3$ units) for $\Upsilon(nS)$ $(n=1,2,3$ respectively) we find them too low, whereas their corresponding values (considering now all contributions) $<O_8(^3S_1)>{\mid}_{tot}$ too high: 480; 220; 160 ( in $10^{-3}$ $GeV^3$ units) respectively, since the COM $\chi_{bJ}(nP)$ contributions are overestimated according to our analysis.         

\ \ \underline{Table 3. Rel. fract.(in \%) for $\Upsilon(1S)$ from CDF}\\
\begin{eqnarray} 
\Upsilon(1S){\mid}_{direct} & = & 50.9 \pm 12.2  \nonumber \\
\Upsilon(2S) + \Upsilon(3S) & = & 11.5 \pm 9.1 \nonumber \\
\chi_{bJ}(1P) & = & 27.1 \pm 8.2 \nonumber \\
\chi_{bJ}(2P) & = & 10.5 \pm 4.6 \nonumber \\
\nonumber
\end{eqnarray}

\ \ \underline{Table 4. Rel.fract.(in \%) for $\Upsilon(1S)$(generation)}\\ 
\begin{eqnarray}
\Upsilon(1S){\mid}_{^3S^{(8)}_1 + CSM} & = & 32.7 + 20.7=52.4 \nonumber \\  
\Upsilon(2S)+\Upsilon(3S){\mid}_{CSM} & = & 4.1 \nonumber \\
\chi_{bJ}(1P){\mid}_{CSM} & = & 25.7 \nonumber \\
\chi_{bJ}(2P){\mid}_{CSM} & = & 16.8 \nonumber  \\
\nonumber
\end{eqnarray}

\section{EXTRAPOLATION TO LHC}

 We predict the differential and integrated cross section at the LHC taking the $<O_8(^3S_1)>$ matrix elements obtained from our fit to CDF data. We have found $gg-gq-q\-\overline{q}$ contributions at LHC energies to be $77\%, 22\%, 1\%$ respectively. Figures 3 and 4  show the differential cross section at LHC energies for $\Upsilon(1S)$ and $\Upsilon(3S)$ respectively. We can see that $\Upsilon(3S)$ should be the best candidate to check the COM since, even at relatively low $P_t$, the $^3S_1^{(8)}$ contribution becomes dominant. In table 5 we summarize the results of the integrated cross section for each $\Upsilon(nS)$ (1S total and direct production).\\

\begin{figure}[htb!]
\centerline{
\epsfig{figure=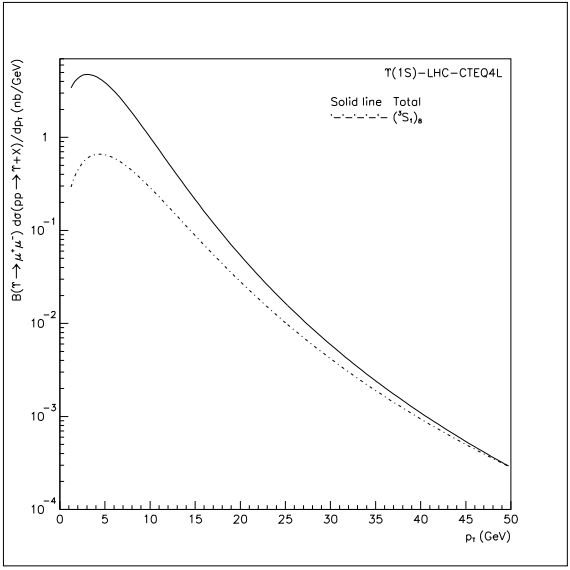,height=7.5cm,width=7.5cm}}
\caption{Theoretical curve using CTEQ4L PDF in $\Upsilon (1S)$ with AP evolution.}
\end{figure}

\begin{figure}[htb!]
\centerline{
\epsfig{figure=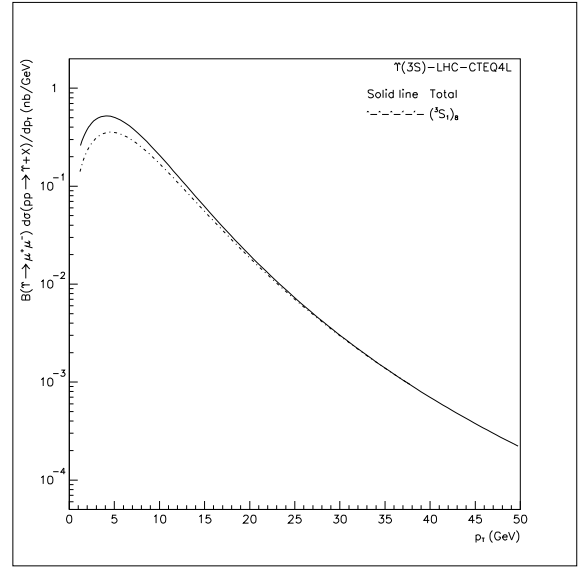,height=7.5cm,width=7.5cm}}
\caption{Theoretical curve using CTEQ4L PDF in $\Upsilon (3S)$ with AP evolution.}
\end{figure}

\ \ \underline{Table 5. Integrated cross sections (nb)}\\
\begin{eqnarray}
\Upsilon(1S){\mid}_{tot}: \sigma & = & 33.4  \nonumber \\  
\Upsilon(1S){\mid}_{dir}: \sigma & = & 12.4  \nonumber \\
\Upsilon(2S){\mid}_{tot}: \sigma & = & 9.4   \nonumber \\
\Upsilon(3S){\mid}_{tot}: \sigma & = & 4.6   \nonumber \\
\nonumber
\end{eqnarray}

\section{CONCLUSIONS}
\begin{itemize}

\item  We have analyzed $\Upsilon(nS)$ hadroproduction at the Tevatron concluding that:\\
 $<O_8^{\Upsilon(nS)}(^3S_1)>$ slightly varies in changing from CTEQ2L to CTEQ4L ( as expected since  both PDF's have a different behaviour, especially at low x) and are compatible with NRQCD velocity scaling rules. We find the $\chi_{bJ}(nP){\mid}_{COM}$  contribution very small. Thereby comparing with Cho and Leibovich work, all $<O_8(^3S_1)>$ values for direct production have to be enhanced.

%\newpage
%.\newpage
%.\newpage

%\begin{figure}[htb!]
%\centerline{
%\epsfig{figure=upsi14_c4l.eps,height=7.5cm,width=7.5cm}}
%\caption{Theoretical curve using CTEQ4L PDF in $\Upsilon (1S)$ with AP evolution.}
%\end{figure}

%values for direct production have to be enhanced.

\item $M_5=5{\times}\biggl(\frac{<O_8(^3P_J)>}{m_b^2}+\frac{<O_8(^1S_0)>}{5}\biggr)$ is compatible with zero.

\item  We have extrapolated to LHC energy providing theoretical curves for the differential transverse momentum distribution and the total cross section for each $\Upsilon(nS)$. Production rates from low to high $P_t$ will afford very interesting physics for bottomonium at the LHC, especially polarisation for the $\Upsilon(3S)$ resonance at $p_t > 20$ GeV. 

\end{itemize}

\subsection*{Acknowledgments}

We acknowledge the B physics working subgroup of the ATLAS collaboration and especially S. Baranov, N. Ellis, M. Kraemer and M. Mangano, for comments and valuable discussions.

%\begin{figure}[htb!]
%\centerline{
%\epsfig{figure=upsi3s14_c4l.eps,height=7.5cm,width=7.5cm}}
%\caption{Theoretical curve using CTEQ4L PDF in $\Upsilon (3S)$ with AP evolution.}
%\end{figure}


\thebibliography{References}
\bibitem{pythia} T. Sj\"{o}strand, Comp. Phys. Comm. {\bf 82} (1994) 74.
\bibitem{nos00} J.L. Domenech and M.A. Sanchis-Lozano, Phys. Lett. {\bf B476} (2000) 65, hep-ph/9911332.
\bibitem{cteq2l} CTEQ Collaboration, Phys. Rev. {\bf D51} (1995) 4763, hep-ph/9410404.
\bibitem{cteq4l} CTEQ Collaboration, Phys. Rev. {\bf D55} (1997) 1280, hep-ph/9606399.
\bibitem{mas0} B. Cano-Coloma and M.A. Sanchis-Lozano, Nucl. Phys. {\bf B508} (1997) 753, hep-ph/9706270. 
\bibitem{mas1} M.A. Sanchis-Lozano, Nucl. Phys. B (Proc. Suppl.) {\bf 86} (2000)
543, hep-ph/9907497 .
\bibitem{eichten} E.J. Eichten and C. Quigg, Phys. Rev. {\bf D52} (1995) 1726.
\bibitem{pdg} C. Caso {\em et al.}, Particle Data Group, EPJ {\bf C3} (1998) 1.
\bibitem{fermi1} G.Feild {\em et al.}, CDF Report/5027.
\bibitem{fermi2} CDF Coll., Phys. Rev. Lett. {\bf 84} (2000) 2094. 
\bibitem{cho} P. Cho and A.K. Leibovich, Phys. Rev. {\bf D53} (1996) 6203.
\end{document}